\documentclass[draft]{agujournal2019}
\usepackage{amsmath}
\usepackage{url} %this package should fix any errors with URLs in refs.
\usepackage{lineno}
\usepackage[inline]{trackchanges} %for better track changes. finalnew option will compile document with changes incorporated.
\usepackage{soul}
\draftfalse

\journalname{Geophysical Research Letters}

\begin{document}

%% ------------------------------------------------------------------------ %%
%  Title
%
% (A title should be specific, informative, and brief. Use
% abbreviations only if they are defined in the abstract. Titles that
% start with general keywords then specific terms are optimized in
% searches)
%
%% ------------------------------------------------------------------------ %%

% Example: \title{This is a test title}

%\title{The effect of the vapor buoyancy on the African easterly jet}
\title{Water vapor buoyancy and the African easterly jet}

%% ------------------------------------------------------------------------ %%
%
%  AUTHORS AND AFFILIATIONS
%
%% ------------------------------------------------------------------------ %%

% Authors are individuals who have significantly contributed to the
% research and preparation of the article. Group authors are allowed, if
% each author in the group is separately identified in an appendix.)

% List authors by first name or initial followed by last name and
% separated by commas. Use \affil{} to number affiliations, and
% \thanks{} for author notes.
% Additional author notes should be indicated with \thanks{} (for
% example, for current addresses).

% Example: \authors{A. B. Author\affil{1}\thanks{Current address, Antartica}, B. C. Author\affil{2,3}, and D. E.
% Author\affil{3,4}\thanks{Also funded by Monsanto.}}

\authors{Heng Quan \affil{1}, Da Yang \affil{2}, William Boos \affil{3}, Tiffany Shaw \affil{4}, Huazhi Ge \affil{5}, Yaoxuan Zeng \affil{4}, Carly KleinStern \affil{4}}

\affiliation{1}{Program in Atmospheric and Oceanic Sciences, Princeton University, Princeton, NJ, US}
\affiliation{2}{Department of Geophysics, Stanford University, Stanford, CA, US}
\affiliation{3}{Department of Earth and Planetary Science, University of California, Berkeley, Berkeley, CA, US}
\affiliation{4}{Department of the Geophysical Sciences, The University of Chicago, Chicago, IL, US}
\affiliation{5}{Division of Geological and Planetary Sciences, California Institute of Technology, Pasadena, CA, US}

%\affiliation{=number=}{=Affiliation Address=}
%(repeat as many times as is necessary)

%% Corresponding Author:
% Corresponding author mailing address and e-mail address:

% (include name and email addresses of the corresponding author.  More
% than one corresponding author is allowed in this LaTeX file and for
% publication; but only one corresponding author is allowed in our
% editorial system.)

% Example: \correspondingauthor{First and Last Name}{email@address.edu}

\correspondingauthor{Heng Quan; Da Yang}{hengquan@princeton.edu; dayang@stanford.edu}

%% Keypoints, final entry on title page.

%  List up to three key points (at least one is required)
%  Key Points summarize the main points and conclusions of the article
%  Each must be 140 characters or fewer with no special characters or punctuation and must be complete sentences

% Example:
% \begin{keypoints}
% \item	List up to three key points (at least one is required)
% \item	Key Points summarize the main points and conclusions of the article
% \item	Each must be 140 characters or fewer with no special characters or punctuation and must be complete sentences
% \end{keypoints}

\begin{keypoints}
% \item Vapor buoyancy from the meridional moisture gradient diagnostically accounts for a 30\% reduction of the African easterly jet magnitude.
% \item Vapor buoyancy offsets 30\% of the meridional temperature gradient contribution to thermal wind balance in the African easterly jet

\item Vapor buoyancy offsets 30\% of the meridional temperature gradient contribution to the peak magnitude of the African easterly jet

\item Two leading GCMs neglect vapor buoyancy and incorrectly have the African easterly jet balanced only by the meridional temperature gradient

% \item The effect of the vapor buoyancy on the African easterly jet becomes stronger under global warming.
% \item Compensation of the meridional temperature gradient by vapor buoyancy increases in climate model projections of global warming
\item Vapor buoyancy is projected to offset more meridional temperature gradient in a warmer climate due to increasing moisture gradient

\end{keypoints}

%% ------------------------------------------------------------------------ %%
%
%  ABSTRACT and PLAIN LANGUAGE SUMMARY
%
% A good Abstract will begin with a short description of the problem
% being addressed, briefly describe the new data or analyses, then
% briefly states the main conclusion(s) and how they are supported and
% uncertainties.

% The Plain Language Summary should be written for a broad audience,
% including journalists and the science-interested public, that will not have 
% a background in your field.
%
% A Plain Language Summary is required in GRL, JGR: Planets, JGR: Biogeosciences,
% JGR: Oceans, G-Cubed, Reviews of Geophysics, and JAMES.
% see http://sharingscience.agu.org/creating-plain-language-summary/)
%
%% ------------------------------------------------------------------------ %%

%% \begin{abstract} starts the second page

%% Abstract less than 150 words!

\begin{abstract}
The African easterly jet (AEJ) is a prominent circulation feature in the tropical atmosphere. It transports mineral dust and generates easterly waves that serve as seeds for hurricanes. Conventional wisdom holds that the AEJ is in thermal wind balance with the positive meridional temperature gradient over North Africa. Here, using reanalysis data, we show that the negative meridional moisture gradient substantially counteracts the effect of the temperature gradient on density in that balance, diagnostically accounting for a 30\% reduction of the AEJ magnitude. Using CMIP6 data, we further show that this effect of vapor buoyancy on the AEJ strengthens under global warming, highlighting the critical role of the spatial distribution of moisture on large-scale circulation. Analysis of the AEJ in CMIP6 models confirms that some models do not include vapor buoyancy in their governing equations, raising questions about the relevance of their projections of climate change in that region.

% Moist air is lighter than dry air at the same temperature and pressure because of a smaller molecular weight. 
% % Recent studies showed that this vapor buoyancy effect significantly affects Earth's radiation budget. 
% The impact of this well-documented vapor buoyancy effect on large-scale atmospheric circulation has been overlooked. Here, we show that vapor buoyancy significantly decelerates the African easterly jet (AEJ) in the northern hemisphere summer. In present climate, the negative moisture (vapor buoyancy) gradient in the lower troposphere makes the AEJ at 600 hPa 30\% slower than the jet induced by the positive temperature (thermal buoyancy) gradient alone. While most climate models consider this effect, a few models neglect the vapor buoyancy and incorrectly attribute the AEJ solely to the meridional temperature gradient. In a warmer climate, a larger moisture gradient offsets 50\% of the AEJ strengthening caused by a larger temperature gradient in global warming simulations, which we quantitatively explain by the Clausius–Clapeyron scaling.    
\end{abstract}

% vapor buoyancy... effect on African easterly jet

% ... meridional buoyancy gradient ... T and q gradient

% while most GCM good, some fail

% effect increases with global warming

%% Plain language summary less than 200 words!

\section*{Plain Language Summary}

The African easterly jet (AEJ) is a strong westward wind maximum about 4000 meters above sea level that forms over northern Africa in summer. It transports mineral dust and affects hurricanes in the tropical Atlantic. Previous studies treated the AEJ as being in a dynamical balance with density gradients created by the north-south atmospheric temperature contrast between the Sahara desert and cooler equatorial regions. Here, using historical data, we demonstrate that the north-south humidity contrast between the dry Sahara and moist Sahel substantially counteracts the effect of the temperature contrast on air density, due to the lower density of water vapor compared to dry air (i.e., vapor buoyancy). The humidity contrast offsets about 30\% of the effect of the temperature gradient, reducing the magnitude of the AEJ needed to maintain dynamical balance. The effect of the humidity contrast on the AEJ strengthens in climate model projections of global warming. Finally, we show that two widely-used climate models do not represent the effect of the humidity contrast on the AEJ, raising questions about the relevance of their projections of climate change in that region.

\section{Introduction}

\begin{figure}
    \centering
    \includegraphics[width=\linewidth]{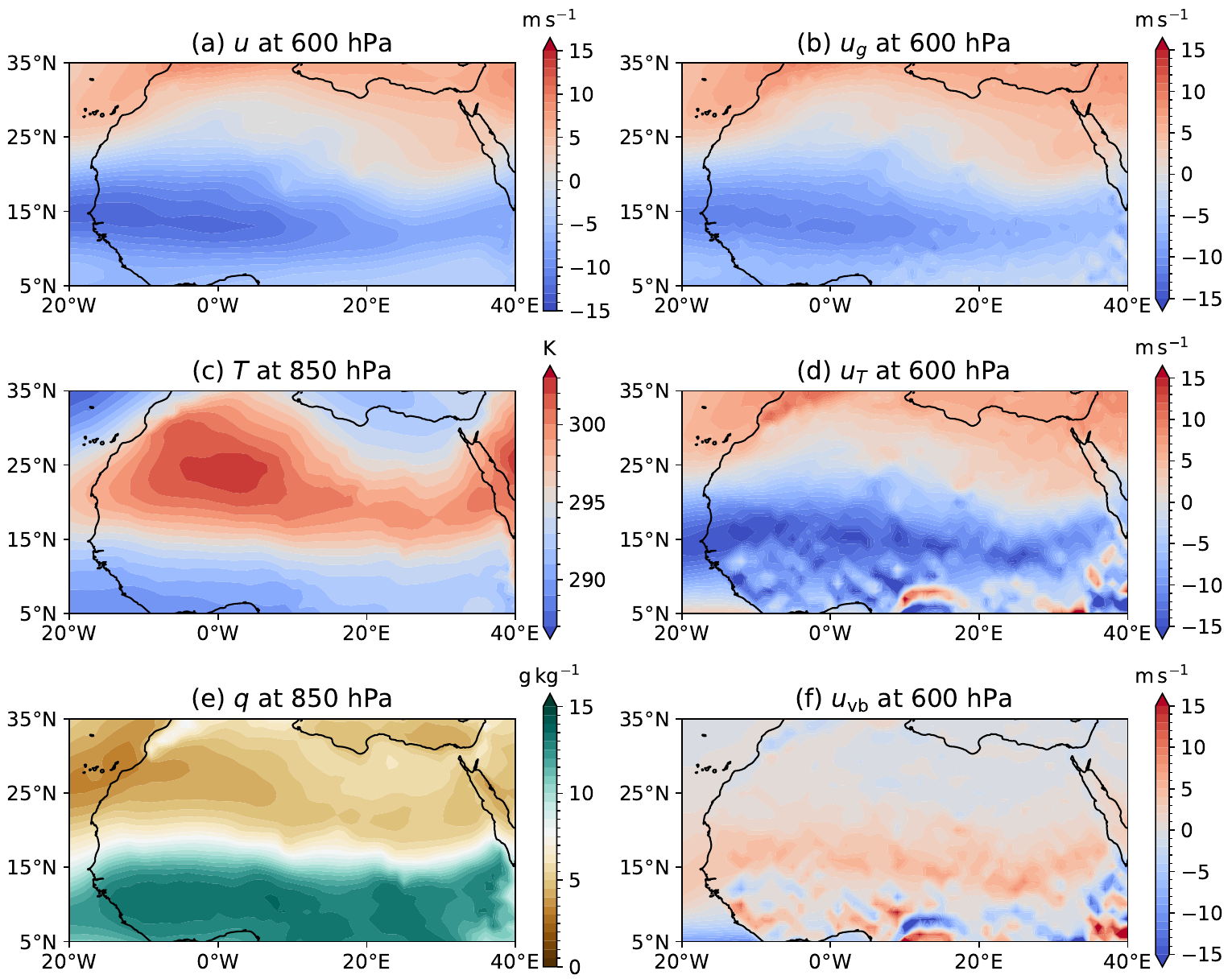}
    \caption{
    (a) Zonal velocity at 600\,hPa, (b) geostrophic zonal velocity ($u_g$) at 600\,hPa, (c) temperature at 850\,hPa, (d) the contribution of the meridional temperature gradient on the geostrophic zonal wind ($u_T$) at 600\,hPa, (e) specific humidity at 850\,hPa and (f) the contribution of the meridional moisture gradient on the geostrophic zonal wind ($u_{\mathrm{vb}}$) at 600\,hPa over northern Africa. All results are ERA5 climatological averages in JJA from 1979 to 2025. See equation \ref{equ:ug_decomposition} for the definition of $u_g$, $u_T$ and $u_{\mathrm{vb}}$.
    }
    \label{fig:AEJ_ERA5_xy}
\end{figure}

The African easterly jet (AEJ) is a prominent circulation feature that forms over northern Africa in summer \cite{cook1999}. Figure \ref{fig:AEJ_ERA5_xy}(a) shows the climatological mean zonal wind at the 600\,hPa level (where the jet is the strongest) in JJA from the ERA5 reanalysis, in which the easterly jet has a maximum velocity around 15\,$\mathrm{m\,s^{-1}}$ near $15^\circ$\,N on the west coast. The cross section in Figure S1 shows that the AEJ is above the low-level westerlies and has a maximum velocity at 600\,hPa. The AEJ transports mineral dust \cite{van2018,bercos2020} and generates easterly waves through barotropic and baroclinic instability \cite{burpee1972,reed1977,pytharoulis1999}, which can affect precipitation in West Africa \cite{fink2003,lubis2015} and serve as precursors for hurricanes in the tropical Atlantic \cite{russell2017,landsea1993,lawton2025}. Therefore, it is crucial to quantitatively understand what controls the magnitude of the AEJ and its response to global warming \cite{skinner2014,patricola2010,patricola2011,bercos2021,akinsanola2025}.

The AEJ is well-known to be in thermal wind balance with the poleward temperature gradient over northern Africa, but the contribution of water vapor buoyancy to that balance has not been assessed. Specifically, high insolation and weak surface latent cooling over the dry Sahara (around $28^\circ$\,N) relative to the intertropical convergence zone (ITCZ, around $7^\circ$\,N) result in a positive meridional surface temperature gradient in summer, which causes a meridional contrast of sensible heat flux and drives the AEJ \cite{cook1999}. Vertical wind shear due to the positive meridional buoyancy gradient in the lower troposphere dominates the low-level westerlies, so the geostrophic zonal wind becomes easterly at 600\,hPa following the thermal wind relation (Figure S1). The AEJ is approximately geostrophic (Figure \ref{fig:AEJ_ERA5_xy}(a)(b)), and previous studies have mostly focused on the balance between the geostrophic zonal wind and the meridional gradient of thermal buoyancy (i.e., temperature) in the lower troposphere (Figure \ref{fig:AEJ_ERA5_xy}(c)). However, buoyancy depends on water vapor content as well as temperature, because humid air is lighter than dry air at the same temperature and pressure due to water's smaller molecular weight \cite{emanuel1994book}. So the negative meridional moisture gradient in the lower troposphere (Figure \ref{fig:AEJ_ERA5_xy}(e)) counteracts the positive temperature gradient in setting the buoyancy gradient, potentially affecting the magnitude of the AEJ.

This study aims to quantitatively assess the effect of vapor buoyancy on the AEJ in both the present climate and a warmer climate, and evaluate whether widely used climate models incorporate this effect. The paper is organized as follows. Section \ref{sec:methods} describes the data used in this study. Section \ref{sec:results} shows that vapor buoyancy with a meridional gradient opposite to that of thermal buoyancy diagnostically accounts for a 30\% reduction of the peak magnitude of the AEJ, and this effect becomes stronger under global warming. This section also identifies two climate models that neglect vapor buoyancy and incorrectly have the AEJ balanced only by the meridional temperature gradient. Finally, Section \ref{sec:conclusions} summarizes the results and conclusions, and discusses implications.

\section{Methods}
\label{sec:methods}

\subsection{Data for the present climate}

We use ERA5 \cite{ERA5} monthly mean reanalysis data on pressure levels and single levels, and calculate averages in JJA from 1979 to 2025. We use monthly outputs from preindustrial control simulations of multiple models in CMIP6, including BCC-CSM2-MR, CAMS-CSM1-0, CESM2, CNRM-CM6-1, GFDL-CM4, GISS-E2-1-G, IITM-ESM, IPSL-CM6A-LR and MIROC6, and calculate the average in JJA over at least 10 years. We also use a hybrid AI-dynamical model, NeuralGCM \cite{kochkov2024}, and calculate the average in JJA from 2000 to 2019 in an AMIP-like simulation provided by \citeA{zhang2025ai}. 

% \subsection{GCM global warming simulations}

% We use the Geophysical Fluid Dynamics Laboratory (GFDL) Forecast-oriented Low Ocean Resolution version of CM2.5 (CM2.5-FLOR) \cite{delworth2012,vecchi2014} to conduct global warming simulations. CM2.5-FLOR is a coupled climate model, whose atmosphere and land components use a horizontal resolution of $0.25^{\circ} \times 0.25^{\circ}$ and 32 vertical levels, and the ocean and sea ice components use lower resolution. The greenhouse gas concentrations except CO$_2$ and aerosol emissions correspond to the conditions of the year 2000. We run the following experiments:

% \begin{enumerate}
%     \item Control simulation. The CO$_2$ concentration is fixed at the value of the year 2000, and the simulation lasts 140 years. Average fields computed from JJA in the last 10 years are referred to as the ``present climate".
%     \item Abrupt 4xCO$_2$ simulation. The model is initialized by the steady state of the control simulation, and the CO$_2$ concentration is abruptly quadrupled upon initialization. The simulation lasts 200 years. Average fields computed from JJA in the last 10 years are referred to as the ``warmer climate".
%     \item +1pct CO$_2$ simulation. The model is initialized by the steady state of the control simulation, and the CO$_2$ concentration increases by 1\% per year for 140 years (a quadrupling by the year 140). Average fields computed from JJA in the last 10 years are referred to as the ``warmer climate", which follows \citeA{quan2025}.
% \end{enumerate}

\subsection{Data for a warmer climate}
We use monthly outputs from abrupt 4xCO$_2$ simulations of multiple models in CMIP6, including BCC-CSM2-MR, CAMS-CSM1-0, CESM2, CNRM-CM6-1, GFDL-CM4, GISS-E2-1-G, IITM-ESM, IPSL-CM6A-LR and MIROC6, and calculate the average in JJA over 10 years no earlier than the simulation years 140-150. Below we show the difference between the ``warmer climate" and ``present climate" (denoted by $\Delta$) normalized by the global mean surface temperature increase (i.e., per 1\,K global mean surface warming) to account for the spread of climate sensitivity across models.

\section{Results}
\label{sec:results}

\subsection{Effect of vapor buoyancy on the AEJ in the present climate}

\begin{figure}
    \centering
    \includegraphics[width=\linewidth]{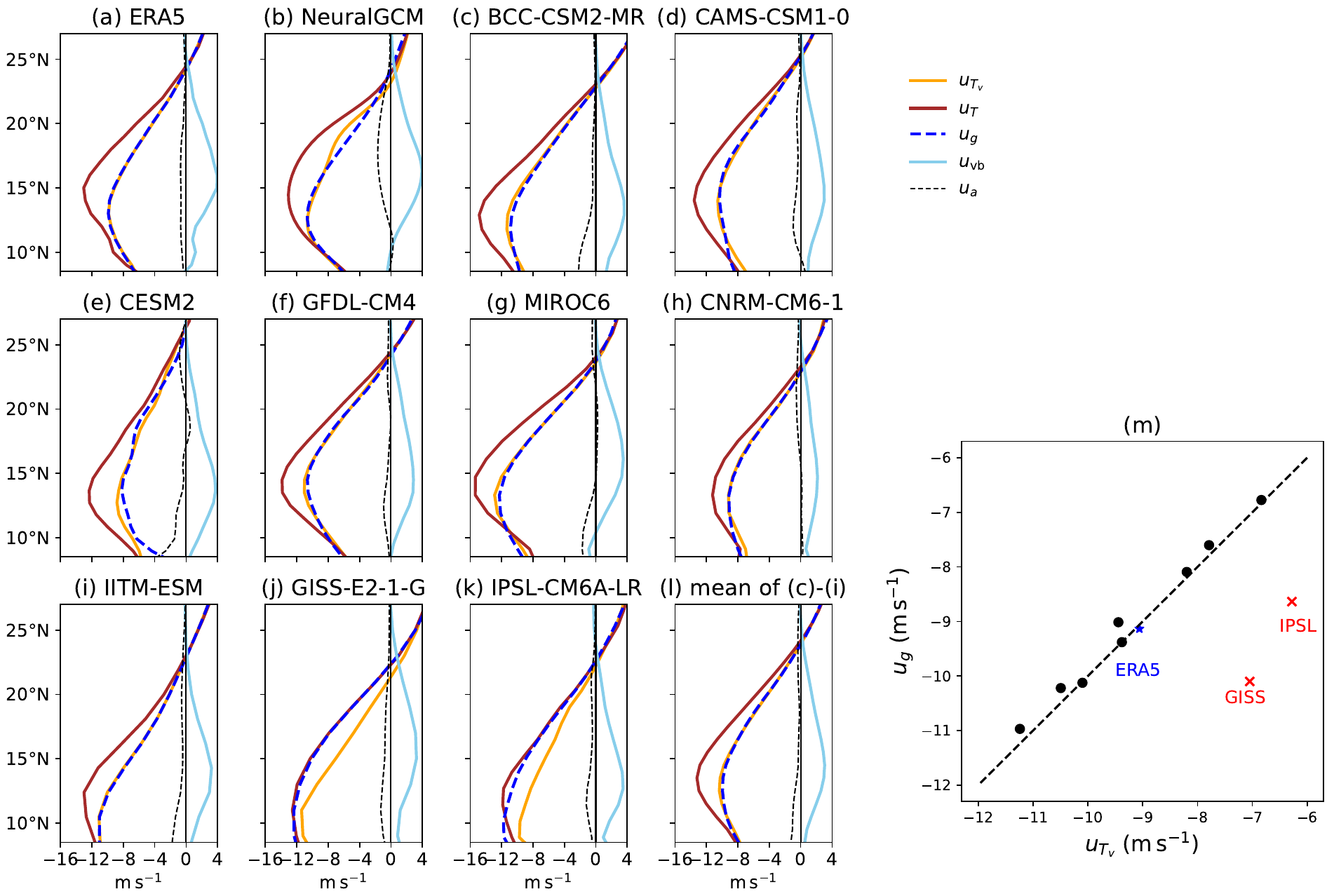}
    \caption{
    (a) - (k) Four zonal velocity components in equation \ref{equ:ug_decomposition} and the ageostrophic zonal velocity ($u_a = u - u_g$ where $u$ is the actual zonal velocity) at 600\,hPa averaged from $20^\circ$\,W to $30^\circ$\,E for ERA5, NeuralGCM (AMIP-like simulation) and CMIP6 preindustrial simulations. All results are climatological averages in JJA over at least 10 years (see Methods). 
    (l) Multi-model-mean zonal velocity profiles for 7 models in (c) - (i). 
    (m) $u_g$ at $15^\circ$\,N against $u_{T_v}$ at $15^\circ$\,N from (a) - (k). 
    Figure S4 and Figure S5 show similar results at 500\,hPa and 700\,hPa.
    }
    \label{fig:u_decomposition}
\end{figure}

We focus on the 600\,hPa level in the main text because the AEJ is the strongest at that level (Figure S1), and we show the vertical structure of the AEJ in supplementary figures. We first study the effect of vapor buoyancy on the AEJ in the present climate. The geostrophic zonal wind $u_g$ in Figure \ref{fig:AEJ_ERA5_xy}(b) is highly similar to the actual zonal wind $u$ in Figure \ref{fig:AEJ_ERA5_xy}(a), and the ageostrophic velocity $u_a$ is an order of magnitude smaller than $u_g$ in ERA5 and all climate models (Figure \ref{fig:u_decomposition}). These results confirm that the AEJ is very nearly in geostrophic balance. We then relate the geostrophic zonal wind at the $p = 600\,\mathrm{hPa}$ level to the meridional buoyancy gradient below that level according to   
\begin{equation}
    \begin{split}
        u_g &= -\frac{1}{fa} \frac{\partial \phi}{\partial \theta} = \underbrace{-\frac{1}{fa} \frac{\partial}{\partial \theta} \left( \phi_s + \int_p^{p_s}\frac{R_d T_v}{p'} \mathrm{d}p' \right)}_{u_{T_v}} \\
        &= \underbrace{-\frac{1}{fa} \frac{\partial}{\partial \theta} \left( \phi_s + \int_p^{p_s}\frac{R_d T}{p'} \mathrm{d}p' \right)}_{u_T} \underbrace{-\frac{1}{fa} \frac{\partial}{\partial \theta} \left( \int_p^{p_s} 0.61\frac{R_d T q}{p'} \mathrm{d}p' \right)}_{u_{\mathrm{vb}}},
    \end{split}
    \label{equ:ug_decomposition}
\end{equation}
where $\theta$ is latitude, $f$ is the Coriolis parameter, $a$ is the radius of Earth, $\phi$ is geopotential, $\phi_s$ is surface geopotential, $p_s$ is surface pressure, $p'$ is pressure and $R_d$ is the gas constant for dry air. In equation \ref{equ:ug_decomposition}, $u_g$ is the geostrophic zonal wind in balance with the meridional pressure gradient, which is nearly identical to $u_{T_v}$ given the high accuracy of hydrostatic balance (Figure \ref{fig:u_decomposition}(a)). Note that we represent buoyancy by the virtual temperature following \citeA{yang2018}, defined as 

\begin{equation}
    \label{equ:Tv}
    T_v = T (1+0.61q),
\end{equation}
where $T$ is temperature and $q$ is specific humidity.

The quantity $u_{T_v}$ depends on the meridional gradient of virtual temperature $T_v$ between the surface and the $p = 600\,\mathrm{hPa}$ level, which we decompose into two components: $u_{T}$ associated with the temperature (thermal buoyancy) gradient and $u_{\mathrm{vb}}$ associated with the moisture (vapor buoyancy) gradient. 
% Note that $u_{T}$ and $u_{\mathrm{vb}}$ are the sums of an atmospheric term ($u_{Ta}$, $u_{\mathrm{vb}a}$) and a surface term ($u_{Ts}$, $u_{\mathrm{vb}s}$):
We express $u_{T}$ as the sum of an atmospheric term $u_{Ta}$ and a surface term $u_{Ts}$:
\begin{equation}
    \begin{split}
        u_{T} = \underbrace{-\frac{1}{fa} \int_p^{p_s}\frac{R_d}{p'} \frac{\partial T}{\partial \theta} \mathrm{d}p'}_{u_{Ta}} + \underbrace{ \left( -\frac{1}{fa} \frac{\partial \phi_s}{\partial \theta} - \frac{1}{fa} \frac{R_d T_{\mathrm{s}}}{p_s} \frac{\partial p_s}{\partial \theta} \right)}_{u_{Ts}}.
    \end{split}
    \label{equ:u_T_decomposition}
\end{equation}
Figure S2 indicates that $u_{Ta}$ (easterly around $15^\circ$\,N), associated with the vertical integral of the meridional temperature gradient, dominates over $u_{Ts}$ (westerly) at $p = 600$\,hPa near $15^\circ$\,N. We incorporate $u_{Ts}$ into $u_T$, so that $u_T$ represents the geostrophic zonal velocity without considering the moisture (vapor buoyancy) gradient.

Similarly, $u_{\mathrm{vb}}$ is also the sum of an atmospheric term $u_{\mathrm{vb}a}$ and a surface term $u_{\mathrm{vb}s}$:
\begin{equation}
    \begin{split}
        u_{\mathrm{vb}} &= \underbrace{-\frac{1}{fa} \int_p^{p_s}\frac{R_d}{p'} \frac{\partial (0.61Tq)}{\partial \theta} \mathrm{d}p'}_{u_{\mathrm{vb}a}} \underbrace{- \frac{1}{fa} \frac{0.61R_d T_sq_s}{p_s} \frac{\partial p_s}{\partial \theta}}_{u_{\mathrm{vb}s}} \\
        & \approx -\frac{1}{fa} \int_p^{p_s}\frac{0.61 R_d T}{p'} \frac{\partial q}{\partial \theta} \mathrm{d}p'.
    \end{split}
    \label{equ:u_vb_decomposition}
\end{equation}
Figure S2 indicates that the surface term $u_{\mathrm{vb}s}$ is comparatively small when zonally averaged over north Africa. Moreover, because $|q \partial T / \partial \theta|$ is an order of magnitude smaller than $|T \partial q / \partial \theta|$ (Figure S3), $u_{\mathrm{vb}}$ at $p = 600$\,hPa is approximately balanced by the term involving the vertical integral of the meridional moisture gradient.

Figure \ref{fig:AEJ_ERA5_xy}(d)(f) shows that $u_T$ is stronger than $u_g$ because the direction of $u_{\mathrm{vb}}$ is opposite to $u_T$ near $15^\circ$\,N. In other words, the negative meridional moisture gradient in the lower troposphere diagnostically accounts for a reduction in the magnitude of the AEJ. The cross section in Figure S1 shows that $u_T$ has its maximum value at the $p = 600\,\mathrm{hPa}$ level, while $u_{\mathrm{vb}}$ increases monotonically with height near $15^\circ$\,N. The small-scale noise in Figure \ref{fig:AEJ_ERA5_xy}(d)(f) is related to the rugged surface, and below we filter this by examining zonal average velocity profiles in northern Africa.

Figure \ref{fig:u_decomposition}(a) shows that the zonal mean geostrophic zonal wind $u_g \approx -10\,\mathrm{m\,s^{-1}}$ at the 600\,hPa level near $15^\circ$\,N is a combination of $u_T \approx -14\,\mathrm{m\,s^{-1}}$ and $u_{\mathrm{vb}} \approx +4\,\mathrm{m\,s^{-1}}$. Therefore, in ERA5 vapor buoyancy from the meridional moisture gradient diagnostically accounts for a 30\% reduction in the peak magnitude of the AEJ near $15^\circ$\,N relative to thermal buoyancy alone. Although vapor buoyancy only accounts for 1\% of the total buoyancy ($0.61q \sim \mathcal{O}(0.01)$ in equation \ref{equ:Tv}), its meridional derivative can be comparable to the meridional derivative of the thermal buoyancy, so vapor buoyancy can be substantial in thermal wind balance of the AEJ.  

Most climate models (Figure \ref{fig:u_decomposition}(c)-(i)) and their multi-model-mean zonal velocity profiles (Figure \ref{fig:u_decomposition}(l)) show similar results to ERA5 (Figure \ref{fig:u_decomposition}(a)). They feature $u_g \approx u_{T_v} = u_T + u_{\mathrm{vb}}$, implying that the AEJ is a combination of $u_T$ associated with the temperature (thermal buoyancy) gradient and $u_{\mathrm{vb}}$ associated with the moisture (vapor buoyancy) gradient. The hybrid AI-dynamical model, NeuralGCM, also shows the same results (Figure \ref{fig:u_decomposition}(b)). These models incorporate vapor buoyancy in their dynamical cores, so they correctly show that vapor buoyancy diagnostically accounts for a reduction of AEJ magnitude (i.e., $|u_{T_v}|<|u_T|$ in Figure \ref{fig:u_decomposition}(b)-(i)).

By contrast, the GISS-E2-1-G model (Figure \ref{fig:u_decomposition}(j)) and the IPSL-CM6A-LR model (Figure \ref{fig:u_decomposition}(k)) feature $u_g \approx u_{T}$, implying that the AEJ is solely balanced by the meridional temperature (thermal buoyancy) gradient. These two models have publicly available source code, and they do not incorporate vapor buoyancy in their dynamical cores to the best of our knowledge \cite{yang2022}. As a result, they lead to an overestimation of the magnitude of the AEJ for a given temperature gradient (i.e., $|u_g|>|u_{T_v}|$ in Figure \ref{fig:u_decomposition}(j)(k)).

The scatter plot in Figure \ref{fig:u_decomposition}(m) clearly shows the contrast between the two classes of models. ERA5 and most models (represented by black dots) are close to the 1:1 line, indicating that $u_g \approx u_{T_v}$ near $15^\circ$\,N with vapor buoyancy correctly accounting for a diagnostic reduction of the AEJ magnitude. The GISS-E2-1-G model and the IPSL-CM6A-LR model (red dots) lie off the 1:1 line, significantly overestimating the magnitude of the AEJ ($|u_g| > |u_{T_v}|$) given their simulated meridional air density gradients, due to neglect of vapor buoyancy for a given temperature gradient. 

\subsection{Stronger effect of the vapor buoyancy on the AEJ in a warmer climate}

Next, we study the changes in vapor buoyancy around the AEJ in a warmer climate. Based on equation \ref{equ:ug_decomposition}, we calculate the change of the geostrophic velocity between a warmer climate and the present climate (details in Methods), denoted by $\Delta$,
\begin{equation}
    \begin{split}
        \Delta u_g &= -\frac{1}{fa} \frac{\partial \Delta \phi}{\partial \theta} = \underbrace{-\frac{1}{fa} \frac{\partial}{\partial \theta} \left( \Delta \phi_s + \Delta \int_p^{p_s}\frac{R_d T_v}{p'} \mathrm{d}p' \right)}_{\Delta u_{T_v}} \\
        &= \underbrace{-\frac{1}{fa} \frac{\partial}{\partial \theta} \left( \Delta \phi_s + \Delta \int_p^{p_s}\frac{R_d T}{p'} \mathrm{d}p' \right)}_{\Delta u_T} \underbrace{-\frac{1}{fa} \frac{\partial}{\partial \theta} \left( \Delta \int_p^{p_s} 0.61\frac{R_d T q}{p'} \mathrm{d}p' \right)}_{\Delta u_{\mathrm{vb}}}.
    \end{split}
    \label{equ:dug_decomposition}
\end{equation}

\begin{figure}
    \centering
    \includegraphics[width=\linewidth]{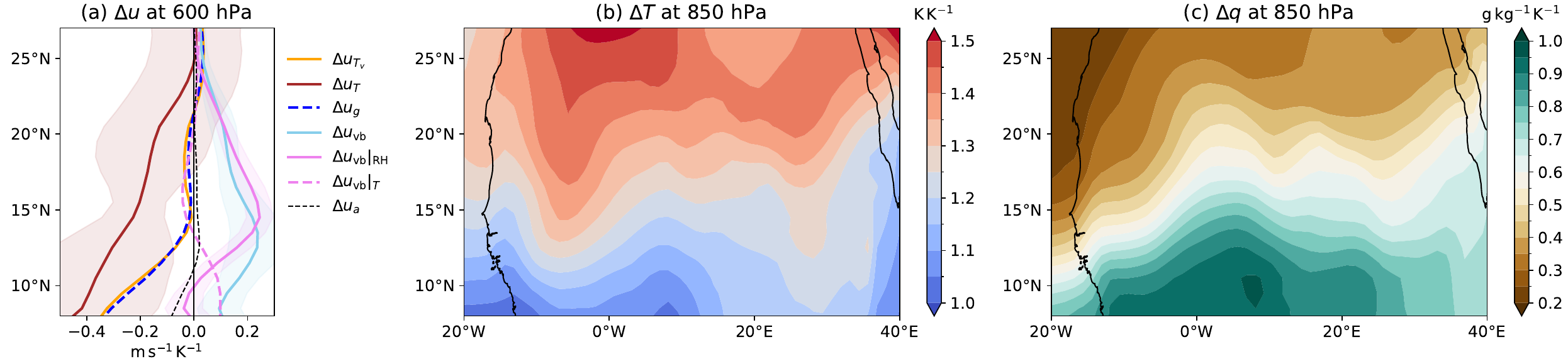}
    \caption{
    Multi-model-mean responses of JJA climatological mean (a) zonal velocity components (as seen in Figure \ref{fig:u_decomposition}) at 600\,hPa averaged from $20^\circ$\,W to $30^\circ$\,E, (b) temperature at 850\,hPa and (c) specific humidity at 850\,hPa normalized by global mean surface temperature increase (i.e., per 1\,K global mean surface warming). The solid and dashed pink curves are two components of $\Delta u_g$: $\Delta u_{\mathrm{vb}}|_{\mathrm{RH}}$ and $\Delta u_{\mathrm{vb}}|_{\mathrm{T}}$ (equation \ref{equ:du_vb_decomposition}). 
    All responses denoted by $\Delta$ are calculated as the difference between CMIP6 abrupt 4xCO$_2$ simulations and preindustrial control simulations (methods). Here we calculate the multi-model-mean responses for 7 models considering vapor buoyancy (excluding GISS-E2-1-G and IPSL-CM6A-LR). The shading in (a) represents +/- one standard deviation across models for $\Delta u_T$, $\Delta u_{\mathrm{vb}}$ and $\Delta u_{\mathrm{vb}}|_{\mathrm{RH}}$.
    Figure S7(a) and Figure S8(a) show the results in (a) at 500\,hPa and 700\,hPa.
    }
    \label{fig:du_CMIP6_warming_mmm_xy}
\end{figure}

We calculate the difference between CMIP6 abrupt 4xCO$_2$ simulations and preindustrial control simulations (see Methods), and show the multi-model-mean responses to global warming for 7 models that represent vapor buoyancy effects (excluding GISS-E2-1-G and IPSL-CM6A-LR). Figure \ref{fig:du_CMIP6_warming_mmm_xy}(a) shows that the AEJ strengthens ($\Delta u_g < 0$) to the south of $15^\circ$\,N and barely changes ($\Delta u_g \approx 0$) to the north of $15^\circ$\,N under global warming at the 600\,hPa level. The southward shift and strengthening of the AEJ with warming relative to the multi-model-mean profile in the present climate (Figure \ref{fig:u_decomposition}(l)) is consistent with that reported in \citeA{kebe2020}. Figure \ref{fig:du_CMIP6_warming_mmm_xy}(a) also shows that $\Delta u_g \approx \Delta u_{T_v} = \Delta u_T + \Delta u_{\mathrm{vb}}$ as suggested by equation \ref{equ:dug_decomposition}, since the 7 models averaged here all incorporate vapor buoyancy in their dynamical cores. Both $u_T$ and $u_{\mathrm{vb}}$ strengthen under global warming: $\Delta u_T < 0$ is associated with an amplified positive meridional temperature (thermal buoyancy) gradient in the lower troposphere (Figure \ref{fig:du_CMIP6_warming_mmm_xy}(b) and Figure S6) \cite{skinner2014,patricola2010,patricola2011,bercos2021,akinsanola2025}, while $\Delta u_{\mathrm{vb}} > 0$ is associated with an amplified negative meridional moisture (vapor buoyancy) gradient in the lower troposphere (Figure \ref{fig:du_CMIP6_warming_mmm_xy}(c) and Figure S6). They offset each other to the north of $15^\circ$\,N, while $\Delta u_T$ dominates to the south of $15^\circ$\,N.

% The actual increase in the magnitude of the AEJ under global warming is much smaller than that due to the larger temperature gradient alone. $\Delta u_g \approx -1\,\mathrm{m\,s^{-1}}$ near $15^\circ$\,N at the 600\,hPa level in Figure \ref{fig:du_CM2.5_warming_xy}(a)(d) is a combination of $\Delta u_T \approx -2\,\mathrm{m\,s^{-1}}$ due to an amplified temperature gradient (Figure \ref{fig:du_CM2.5_warming_xy}(b)(e)) and $\Delta u_{\mathrm{vb}} \approx +1\,\mathrm{m\,s^{-1}}$ due to an amplified moisture gradient (Figure \ref{fig:du_CM2.5_warming_xy}(c)(f)). Therefore, enhanced vapor buoyancy from the increased moisture gradient offsets 50\% of the AEJ strengthening caused by the increased temperature gradient under global warming. 

The magnitude of the AEJ at the 600\,hPa level depends on the near-surface wind and the meridional buoyancy gradient between the surface and the 600\,hPa level. The cross section in Figure S6 shows that both the near-surface westerlies and the vertical wind shear (associated with the lower-tropospheric meridional $T_v$ gradient) strengthen under global warming \cite{bercos2021,akinsanola2025}. They approximately offset each other and result in a near-zero AEJ response at the 600\,hPa level north of $15^\circ$\,N, while the vertical wind shear dominates and results in a stronger AEJ at the 600\,hPa level to the south of $15^\circ$\,N. Above 600\,hPa the AEJ strengthens because the vertical wind shear dominates (Figure S7), while below 600\,hPa the AEJ weakens near $15^\circ$\,N because the surface westerlies dominate (Figure S8).

The profile of $\Delta u_{\mathrm{vb}}$ in Figure \ref{fig:du_CMIP6_warming_mmm_xy}(a) can be quantitatively explained by a Clausius–Clapeyron scaling. The relative changes of $T$ and $p_s$ in equation \ref{equ:u_vb_decomposition} under global warming are negligible (around $1\%$) compared to the relative change of $q$, so the change of $u_{\mathrm{vb}}$ can be expressed as

\begin{equation}
    \begin{split}
        \Delta u_{\mathrm{vb}} & \approx -\frac{1}{fa} \int_p^{p_s}\frac{0.61 R_d T}{p'} \frac{\partial \Delta q}{\partial \theta} \mathrm{d}p' \\
        & \approx \underbrace{-\frac{1}{fa} \int_p^{p_s}\frac{0.61 R_d T}{p'} \frac{\partial (\alpha q \Delta T)}{\partial \theta} \mathrm{d}p'}_{\Delta u_{\mathrm{vb}}|_{\mathrm{RH}}} \underbrace{- \frac{1}{fa} \int_p^{p_s}\frac{0.61 R_d T}{p'} \frac{\partial (q \Delta \ln{\mathrm{RH}})}{\partial \theta} \mathrm{d}p'}_{\Delta u_{\mathrm{vb}}|_{\mathrm{T}}}, 
    \end{split}
    \label{equ:du_vb_decomposition}
\end{equation}
where $\alpha \approx 0.07\, \mathrm{K^{-1}}$ is the rate at which the saturation vapor pressure increases with temperature at the reference climate according to the Clausius-Clapeyron relationship, $\Delta T$ is the change in temperature, and $\Delta \ln{\mathrm{RH}}$ is the relative change in relative humidity. The notation $|_A$ means ``with A fixed". The theoretical estimates of $\Delta u_{\mathrm{vb}}|_{\mathrm{RH}}$ and $\Delta u_{\mathrm{vb}}|_{\mathrm{T}}$ are shown by the pink curves in Figure \ref{fig:du_CMIP6_warming_mmm_xy}(a). The quantity $\Delta u_{\mathrm{vb}}|_{\mathrm{RH}}$ (the change assuming fixed relative humidity) captures the profile of $\Delta u_{\mathrm{vb}}$ and its peak value of $+0.2\,\mathrm{m\,s^{-1}}$ ($+7\%$) per 1\,K global mean surface warming near $15^\circ$\,N, while the term $\Delta u_{\mathrm{vb}}|_{\mathrm{T}}$ (due to changes only in relative humidity) accounts for a slight northward shift of the AEJ near $15^\circ$\,N. $\Delta u_{\mathrm{vb}}$ is overall dominated by $\Delta u_{\mathrm{vb}}|_{\mathrm{RH}}$, but is dominated by $\Delta u_{\mathrm{vb}}|_{\mathrm{T}}$ to the south of $11^\circ$\,N.   

\begin{figure}
    \centering
    \includegraphics[width=\linewidth]{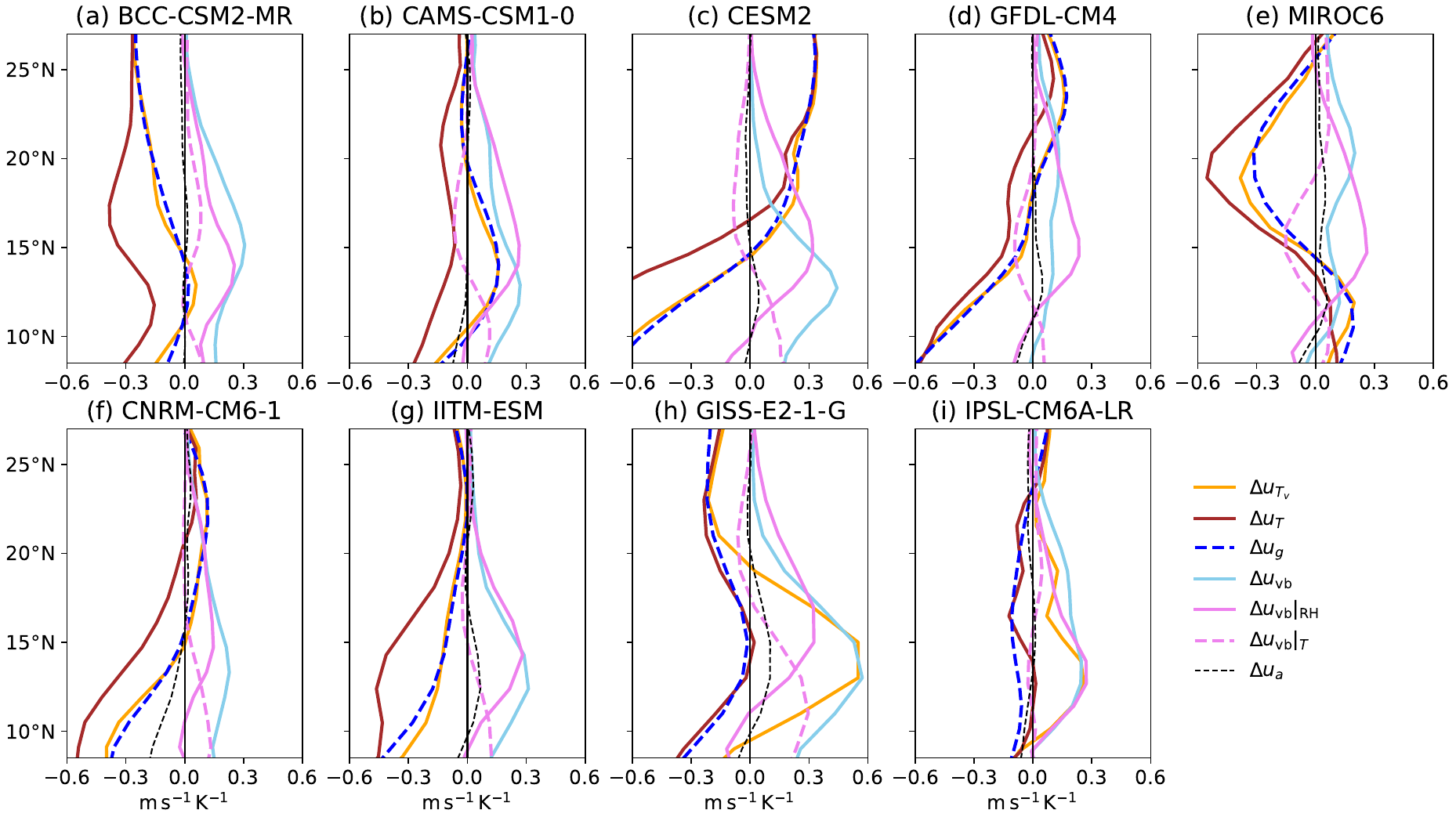}
    \caption{
    Same as Figure \ref{fig:du_CMIP6_warming_mmm_xy}(a) but for individual models.
    Figure S9 and Figure S10 show the results at 500\,hPa and 700\,hPa.
    }
    \label{fig:du_CMIP6_warming_individual_xy}
\end{figure}

We show the responses of the AEJ at the 600\,hPa level to global warming for individual CMIP6 GCMs in Figure \ref{fig:du_CMIP6_warming_individual_xy}. Models exhibit a diverse set of changes in the meridional profile of $\Delta u_g$, with strengthening (a,g,h,i), weakening (b), northward shift (e) and southward shift (c,d,f) all appearing in different models. A recent study by \citeA{kuete2026} found a large inter-model spread of the responses of the AEJ to global warming in autumn (September to November), and here we show a similar variability across models in summer (June to August). For models that represent vapor buoyancy (Figure \ref{fig:du_CMIP6_warming_individual_xy}(a)-(g)), $\Delta u_g \approx \Delta u_{T_v} = \Delta u_T + \Delta u_{\mathrm{vb}}$ and $\Delta u_g$ depends on the relative magnitude of $\Delta u_T$ and $\Delta u_{\mathrm{vb}}$. For models neglecting vapor buoyancy (Figure \ref{fig:du_CMIP6_warming_individual_xy}(h)(i)), $\Delta u_g \approx \Delta u_T$ and is not affected by $\Delta u_{\mathrm{vb}}$. 

Figure \ref{fig:du_CMIP6_warming_individual_xy} and the shading in Figure \ref{fig:du_CMIP6_warming_mmm_xy}(a) both indicate that the large inter-model spread of $\Delta u_g$ is primarily rooted in the inter-model spread of $\Delta u_T$, which features strengthening (a,b,g,h,i), northward shift (e) and southward shift (c,d,f) in different models. Different spatial patterns of $\Delta u_T$ imply different spatial patterns of temperature increase across models in the lower troposphere. By contrast, $\Delta u_{\mathrm{vb}}$ associated with the meridional moisture gradient is more robust (i.e., it has a smaller inter-model spread). All models agree that $\Delta u_{\mathrm{vb}} > 0$, implying a robust amplification of the meridional moisture gradient (as shown by Figure \ref{fig:du_CMIP6_warming_mmm_xy}(c)) across models in the lower troposphere. Furthermore, the component $\Delta u_{\mathrm{vb}}|_{\mathrm{RH}}$ (assuming fixed relative humidity) has a even smaller inter-model spread than $\Delta u_{\mathrm{vb}}$, as $\Delta u_{\mathrm{vb}}|_{\mathrm{RH}}$ is constrained by the Clausius–Clapeyron scaling in individual models. The component $\Delta u_{\mathrm{vb}}|_{\mathrm{T}}$ (due to the change in relative humidity) is less constrained than $\Delta u_{\mathrm{vb}}|_{\mathrm{RH}}$, because relative humidity is affected by large-scale dynamics as well as convection, which differs a lot across models. In most models $\Delta u_{\mathrm{vb}}$ is dominated by $\Delta u_{\mathrm{vb}}|_{\mathrm{RH}}$ near $15^\circ$\,N while dominated by $\Delta u_{\mathrm{vb}}|_{\mathrm{T}}$ to the south of $11^\circ$\,N.

\section{Conclusions and discussions}
\label{sec:conclusions}

Although vapor buoyancy only accounts for about 1\% of the total buoyancy ($0.61q \sim \mathcal{O}(0.01)$ in equation \ref{equ:Tv}), we demonstrate that it can significantly affect the momentum balance of the AEJ. In the present climate, vapor buoyancy diagnostically accounts for a 30\% reduction of the peak magnitude of the AEJ (at 15$^\circ$N and 600 hPa) relative to thermal buoyancy alone. Under global warming, the effect of vapor buoyancy on the AEJ strengthens according to the Clausius–Clapeyron scaling (around $+7\%$ per 1\,K global mean surface warming). While most climate models correctly incorporate the vapor buoyancy effect, the GISS-E2-1-G model and the IPSL-CM6A-LR model neglect the vapor buoyancy effect and incorrectly have the AEJ balanced only by the meridional temperature (thermal) gradient; this could potentially lead to an overestimation of the magnitude of the AEJ if eddy transports of heat or momentum do not compensate. Recent studies found that the vapor buoyancy has profound impacts on convective aggregation, low cloud cover and climate sensitivity \cite{yang2018,yang2020,seidel2020,yang2022}. We demonstrate that it also affects the momentum balance of the large-scale circulation, emphasizing the importance of representing vapor buoyancy in climate models.

% Some previous studies found that the AEJ is stronger in years with a drier Sahel than in years with a wetter Sahel \cite{Newell1984,grist2002,cornforth2009}, which is expected based on our findings. However, previous studies focused on how moisture affects the magnitude of the AEJ by changing the meridional temperature (thermal buoyancy) gradient, which is an indirect effect. We propose that the magnitude of the AEJ may be directly affected by moisture through vapor buoyancy, providing a new perspective to interpret the interannual variability of the AEJ.

Some previous studies found that the AEJ near 15$^\circ$N is stronger in years with a drier Sahel (around 17$^\circ$N) than in years with a wetter Sahel \cite{Newell1984,grist2002}, because a wetter Sahel means a smaller meridional contrast of the surface latent cooling between Sahel and the ITCZ (around 7$^\circ$N), which results in a smaller meridional temperature (thermal buoyancy) gradient in the lower troposphere. These studies focused on the effect of moisture on thermal buoyancy, while here we propose that moisture could also affect the AEJ through vapor buoyancy, providing a new perspective to interpret the interannual variability of the AEJ.

All results in this study are diagnostic in a sense that we study the effect of vapor buoyancy on the AEJ with fixed thermal buoyancy. A next step would be to compare the magnitude of the AEJ in a climate model control simulation against that in a mechanism-denial simulation in which the vapor buoyancy effect is turned off, using a methodology similar to that of \citeA{yang2022}. 
%Such a comparison would indicate how the indirect effect of vapor buoyancy on the AEJ (through changes in the thermal buoyancy) compares to the direct effect that this study focuses on. 
Such a comparison would indicate how the effect of vapor buoyancy on the AEJ through changes in the thermal buoyancy compares to the effect of vapor buoyancy on the AEJ with fixed thermal buoyancy that this study focuses on. Future research on downstream impacts, such as the impacts of vapor buoyancy on dust mineral transport and Atlantic hurricanes through the AEJ, is also warranted.

\section{Open Research}
The ERA5 data used in this study can be found at \url{https://cds.climate.copernicus.eu/datasets/reanalysis-era5-pressure-levels-monthly-means?tab=download}. The CMIP6 output used in this study can be found at \url{https://metagrid.esgf-west.org/search}. The NeuralGCM output used in this study can be found at \url{https://tigress-web.princeton.edu/~bosongz/emulators/}.

% The data derived from CMIP6 data can be found at \url{https://doi.org/10.5281/zenodo.6799004} \cite{andrews2022}. The radiative kernel analysis follows methods provided by Chenggong Wang at \url{https://github.com/ChenggongWang/Radiative_Response_with_Radiative_Kernel}.

% AGU requires an Availability Statement for the underlying data needed to understand, evaluate, and build upon the reported research at the time of peer review and publication.

% Authors should include an Availability Statement for the software that has a significant impact on the research. Details and templates are in the Availability Statement section of the Data and Software for Authors Guidance: \url{https://www.agu.org/Publish-with-AGU/Publish/Author-Resources/Data-and-Software-for-Authors#availability}

% It is important to cite individual datasets in this section and, and they must be included in your bibliography. Please use the type field in your bibtex file to specify the type of data cited. Some options include Dataset, Software, Collection, ComputationalNotebook. Ex: 
% \\
% \begin{verbatim}

% @misc{https://doi.org/10.7283/633e-1497,
%   doi = {10.7283/633E-1497},
%   url = {https://www.unavco.org/data/doi/10.7283/633E-1497},
%   author = {de Zeeuw-van Dalfsen, Elske and Sleeman, Reinoud},
%   title = {KNMI Dutch Antilles GPS Network - SAB1-St_Johns_Saba_NA P.S.},
%   publisher = {UNAVCO, Inc.},
%   year = {2019},
%   type = {dataset}
% }

% \end{verbatim}

% For physical samples, use the IGSN persistent identifier, see the International Geo Sample Numbers section:
% \url{https://www.agu.org/Publish-with-AGU/Publish/Author-Resources/Data-and-Software-for-Authors#IGSN}
% %%%%%%%%%%%%%%%%%%%%%%%%%%%%%%%%%%%%%%%%%%%%%%%

\acknowledgments
The project was made possible by the summer school Rossbypalooza (2024): Climate and Extreme Events. Rossbypalooza is funded by the University Corporation for Atmospheric Research, NSF, and the Department of the Geophysical Sciences and Graduate Council at the University of Chicago. D.Y. is supported by an NSF CAREER Award (AGS-2048268) and a Packard Fellowship in Science and Engineering. 

% This section is optional. Include any Acknowledgments here.
% The acknowledgments should list:\\
% All funding sources related to this work from all authors\\
% Any real or perceived financial conflicts of interests for any author\\
% Other affiliations for any author that may be perceived as having a conflict of interest with respect to the results of this paper.\\
% It is also the appropriate place to thank colleagues and other contributors. AGU does not normally allow dedications.

\bibliography{agusample}

\end{document}